# Herbig Ae Young Star's Infrared Spectrum Identified By Hydrocarbon Pentagon-Hexagon Combined Molecules

NORIO OTA

Graduate School of Pure and Applied Sciences, University of Tsukuba,
1-1-1 Tenoudai Tsukuba-city 305-8571, Japan;　　n-otajitaku@nifty.com

Infrared spectrum (IR) of Herbig Ae young star was reproduced and classified by hydrocarbon pentagon-hexagon combined molecules by the quantum chemical calculation. Observed IR list by B. Acke et al. was categorized to four classes. Among 53 Herbig Ae stars, 26 samples show featured IR pattern named Type-D, which shows common IR peaks at 6.2 8.3, 9.2, 10.0, 11.2, 12.1, and 14.0 micrometer. Typical example is HD144432. Calculation on di-cation molecule $(C_{12}H_8)^{2+}$ having hydrocarbon one pentagon and two hexagons shows best coincidence at 6.1, 8.2, 9.2, 9.9, 11.2, 12.2, and 14.1 micrometer. There are some variations in Type-D. Spectrum of HD37357 was explained by a mixture of di-cation $(C_{12}H_8)^{2+}$ and tri-cation $(C_{12}H_8)^{3+}$. Ubiquitously observed spectrum Type-B was observed in 12 samples of Acke's list. In case of HD85567, observed 16 peaks were precisely reproduced by a single molecule $(C_{23}H_{12})^{2+}$. There is a mixture case with Type-D and Type-B. Typical example was HD142527. In this study, we could identify hidden carrier molecules for all types of infrared spectrum in Herbig Ae stars.

1, INTRODUCTION

Herbig Ae stars are pre-main-sequence stars having few solar masses. Our Sun may have a similar situation in young age. In order to understand how polycyclic aromatic hydrocarbon (PAH) was created in space, infrared spectrum (IR) in Herbig Ae star is very important. If we could identify hidden carrier molecules, we could suggest one inevitable rout of a building block of "life" in space. There is an important observed IR data edited by B. Acke et al. including 53 samples (Acke 2010). In this paper, such spectra were analyzed by the quantum chemical calculation applying model molecules. IR of Herbig Ae stars will be categorized to four classes referencing our previous study (Ota 2017d). In Table 1, type name and typical spectrum pattern was illustrated. Type-A has strong peak at 11.3 micrometer (Kaneda 2007), of which carrier molecule was previously suggested as pure-carbon polycyclic molecule $(C_{23})^{2+}$ (Ota 2017d). Unfortunately, we could not find any example in Acke's list. Type-B is ubiquitously observed in many nebula and galaxy. Well observed wavelength is 6.2, 7.7, 8.6, 11.3, and 12.7 micrometer. In Acke's list, we could find 12 samples. Typical carrier molecule of Type-B was suggested to be di-cation $(C_{23}H_{12})^{2+}$ having hydrocarbon two pentagons coupled with five hexagons (Ota 2014, 2017a, 2017b, 2017c, 2017d). Type-C shows similar spectrum with Type-B, but very strong peak at 11.3 micrometer. In the list, 15 samples were recognized. Previous study discussed that Type-C may be a mixture with Type-B and Type-A, and/or mixture with neutral PAH as like $(C_{24}H_{12})$. Type-D is a majority in Acke's list. There are 26 samples showing complex spectrum patterns.  Featured unusual IR peaks are found at 6.2 8.3, 9.2, 10.0, 11.2, 12.1, and 14.0 micrometer. This paper mainly concerns to discover Type-D hidden carrier molecules by the quantum chemical calculation. In order to find out a molecule, I traced a history of star's death and birth and made one possible hypothesis named "Top down astrochemical evolution model" (Ota 2017a), that is, evolution steps as like (1) nucleation of graphene molecule during supernova expansion, (2) void creation by high speed proton, (3) hydrogenation by low speed proton, and (4) ionization by high energy photon. This basic concept was very useful to find hidden molecules for every type. We can expect to identify Type-D hidden molecule under the same concept. Finally, we will suggest simple cationic molecules $(C_{12}H_8)^{n+}$ (n=1, 2 and 3).

Table 1, Classification of Herbig Ae Star infrared spectrum based on observed data
edited by B. Acke et al., (Acke 2010).

| IR Type | Spectrum<br>Flux vs. Wavelength (micrometer) | Carrier Molecule Candidates | Herbig Ae star's IR List by Acke et al. (Acke 2010) |
|---|---|---|---|
| Type-A | 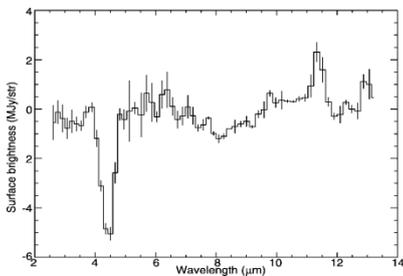<br>NGC1316 (Kaneda 2007) | $(C_{23})^{2+}$<br>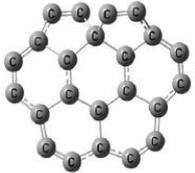<br>(ota 2017d) | **Not found**<br>in Acke's IR List |
| Type-B | 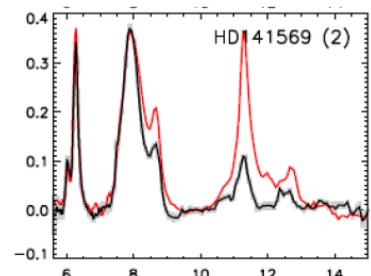<br>Black curve (Acke 2010) | $(C_{23}H_{12})^{2+}$<br>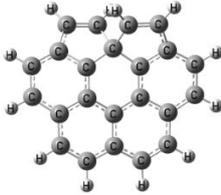<br>(Ota 2014, 2017a) | HD141569, HD34282, HD35187, HD37411, HD85567, HD101412,<br>(12 samples) |
| Type-C | 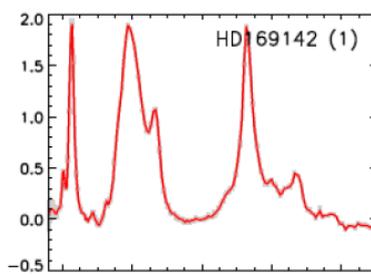<br>Red curve (Acke 2010) | $(C_{23}H_{12})^{2+} + (C_{23})^{2+}$<br>$+ (C_{24}H_{12})$<br>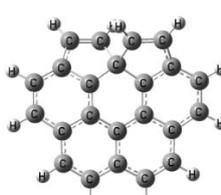<br>(ota 2017d) | HD169142, HD31293, RR Tau, HD72106S, HD97048, HD142527,<br>(15 samples) |
| Type-D | 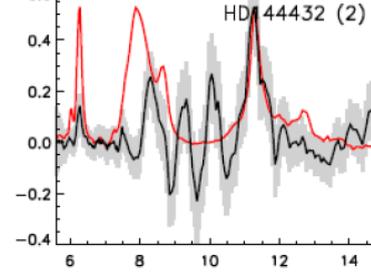<br>Black curve (Acke 2010) | **Not Identified**<br>This paper's model<br>$(C_{12}H_8)^{2+}$<br>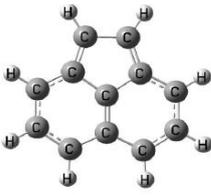 | HD144432, HD31648, HD37258, BF Ori HD37357, HD104237, HD203024<br>(26 samples) |

## 2, TOP DOWN ASTROCHEMICAL EVOLUTION MODEL

In our previous study (Ota 2017a), we discussed "Top down astrochemical evolution model", which resulted to identify Type-B. Typical model molecule is $(C_{23}H_{12})^{2+}$. After that, we added pure carbon polycyclic molecule as like $(C_{23})^{2+}$ (Ota 2017d) identified as Type-A. Here, we like to add smaller graphene as typically illustrated in Figure 1 starting from $(C_{13})$ having pure carbon three hexagons. Interstellar dust was assumed to be created after supernova explosion (Nozawa 2003, 2006). After explosion, graphene molecule would be ejected to surrounding space, which may become a seed for organic dust. Typical model is $(C_{13})$ as shown in (a). Ejected carbon diffuses to surrounding space at a high speed and will collide with interstellar gas, mainly with proton $H^+$. As shown in (b), high speed collision will make a void inside of graphene. It is essential that there occurs quantum mechanical configuration change as shown in (c), where void graphene will be rearranged to $(C_{12})$ having one carbon pentagons combined with two hexagons. After new star born, as like Herbig Ae stars, there accompanies low speed proton irradiation (stellar wind), which may cause hydrogenation as noted in (d) as $(C_{12}H_8)$. At the same time, there occurs photoionization (e) resulting cation molecule as like $(C_{12}H_8)^{n+}$ (n=1, 2, 3….).

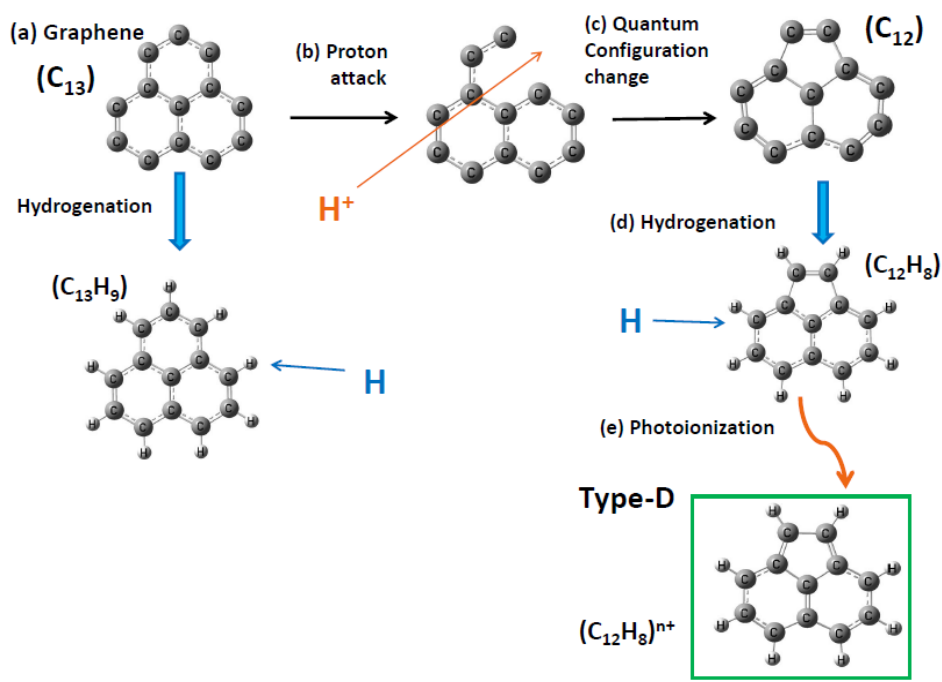

Figure 1, Top down astrochemical evolution mechanism for a model molecule of $(C_{13})$.

## 3, CALCULATION METHOD

In quantum chemistry calculation, we have to obtain total energy, optimized atom configuration, and infrared vibrational mode frequency and strength depend on a given initial atomic configuration, charge and spin state Sz. Density functional theory (DFT) with unrestricted B3LYP functional was applied utilizing Gaussian09 package (Frisch et al. 1984, 2009) employing an atomic orbital 6-31G basis set. The first step calculation is to obtain the self-consistent energy, optimized atomic configuration and spin density. Required convergence on the root mean square density matrix was less than $10^{-8}$ within 128 cycles. Based on such optimized results, harmonic vibrational frequency and strength was calculated. Vibration strength is obtained as molar absorption coefficient ε (km/mol.). Comparing DFT harmonic wavenumber $N_{DFT}$ (cm$^{-1}$) with experimental data, a single scale factor 0.965 was used (Ota 2015). Concerning a redshift for the anharmonic correction, in this paper we did not apply any correction to avoid over estimation in a wide wavelength representation from 2 to 30 micrometer.

Corrected wave number N is obtained simply by N (cm$^{-1}$) = $N_{DFT}$ (cm$^{-1}$) x 0.965.

Wavelength λ is obtained by λ (micrometer) = 10000/N(cm$^{-1}$).

Reproduced IR spectrum was illustrated in a figure by a decomposed Gaussian profile with full
width at half maximum FWHM=4cm$^{-1}$.

## 4, TYPE-D INFRARED SPECTRUM

Majority of Herbig Ae IR was Type-D. Typical example was illustrated on top of Figure 2 (example 1) from observed IR of HD144432 reported by B. Acke et al. (Acke 2010). As illustrated by a black curve, we can see main peaks at 6.2 8.3, 9.2, 10.0, 11.3, 12.1, and 14.0 micrometer. For comparison, typical Type-B spectrum is noted by a red curve (HD169142), which is ubiquitously observed in many nebula and galaxy. Type-D is very different with –B. In order to find out similar IR characteristics by the quantum chemical calculation, many model molecules was tested. Among them, good examples were obtained from $(C_{12}H_8)^{n+}$ series. Figure 3 shows calculated results of $(C_{12}H_8)^{n+}$ on n=0 (neutral), +1 (mono cation), +2, +3, +4, and n=-1(anion). Comparing with those calculated spectra with HD144432, best coincident one was di-cation $(C_{12}H_8)^{2+}$. Calculated wavelength was 6.1, 8.2, 9.2, 9.9, 11.3, 12.2, and 14.1 micrometer. It should be noted that calculated wavelength could reproduce observed one by an accuracy within +/- 0.1 micrometer. There remains a problem that calculated peak height at 6.2, 6.8, and 7.1 micrometer are higher than observed one. Relative strength between those three peaks is similar for both observation and calculation. On a side of observation, detector window responsibility may be a problem in a range of 5~7 micrometer. On calculation, anharmonic molecular vibration on 6.2 micrometer should be considered in detail.

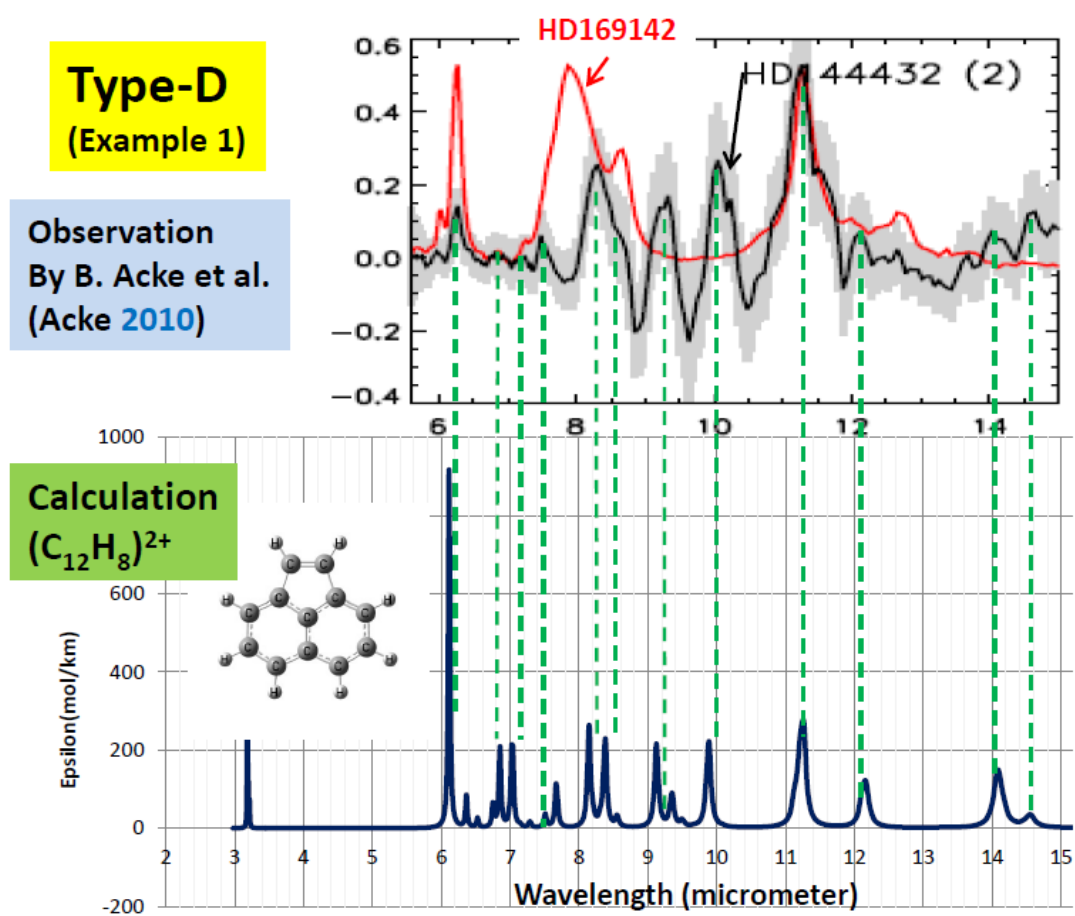

Figure 2, Infrared spectrum of Type-D. Typical spectrum is illustrated on top from observation on HD144432 among a list by B. Acke et al. (Acke 2010). Down panel show calculated result of di-cation molecule $(C_{12}H_8)^{2+}$, which shows best coincidence with observation.

Another Type-D spectrum was shown in Figure 4. Observation sample was HD37357 as example 2 (Acke 2010). Major peaks are similar with example 1 as 6.2, 8.2, 9.3, 10.0, 11.2, 12.1, and 14.0 micrometer. However, we can see additional peaks at 7.5, 7.8, 10.5, 12.6, and 14.6 micrometer. Comparing calculated results in Figure3, we noticed that tri-cation (C12H8)3+ can satisfy such additional peaks. Figure 4 show parallel view between HD37357, $(C_{12}H_8)^{2+}$, and $(C_{12}H_8)^{3+}$. By a suitable mixture with two different ionized spectrum, we can expect to reproduce observed IR of

HD37357.

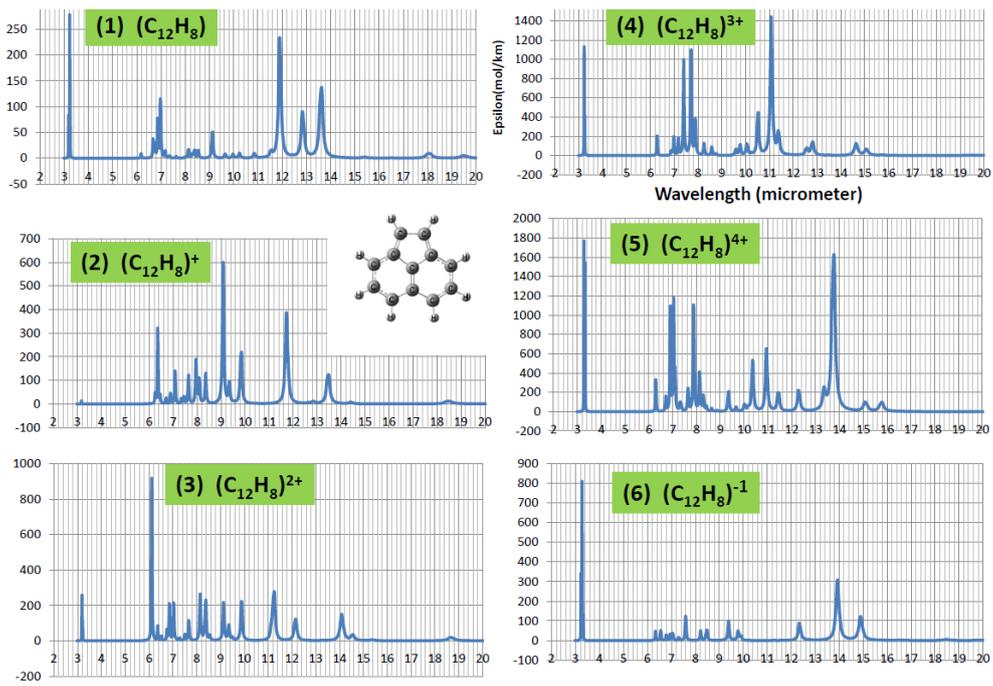

Figure 3, Calculated spectra of $(C_{12}H_8)^{n+}$ molecules.

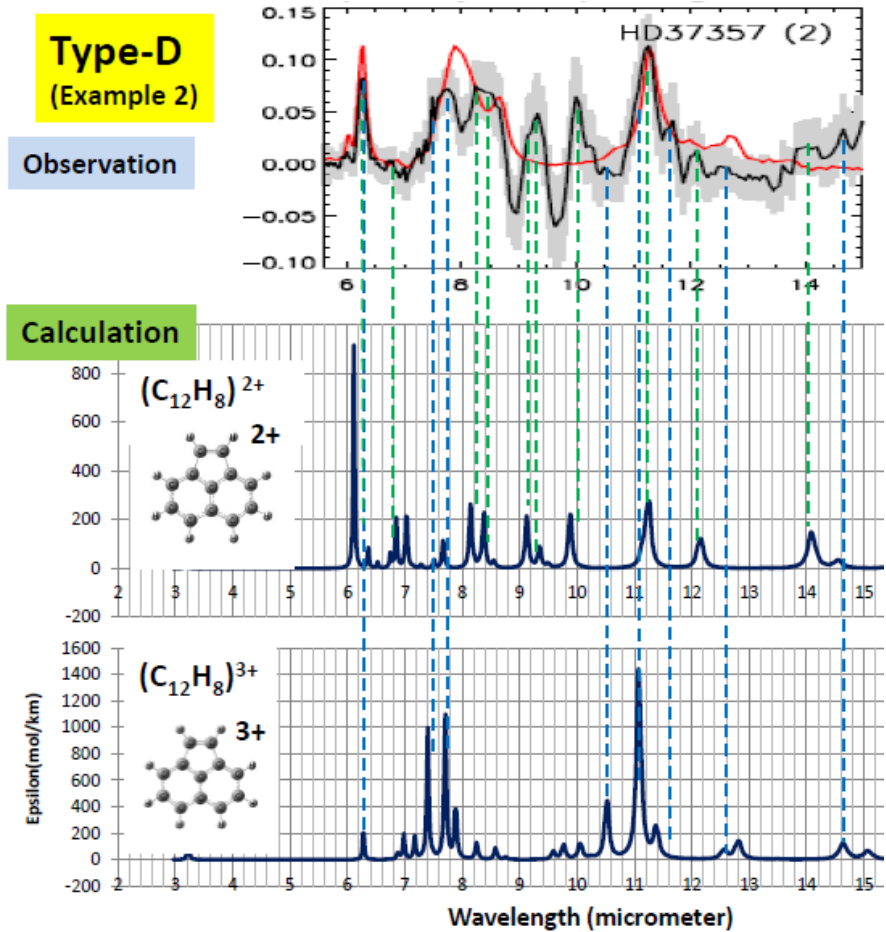

Figure 4, Observed spectrum of HD37357 could be reproduced by a mixture of di-cation $(C_{12}H_8)^{2+}$ and tri-cation $(C_{12}H_8)^{3+}$.

5, MIXED TYPE-B AND Type-D INFRARED SPECTRUM

Type-B was well reproduced by a molecule $(C_{23}H_{12})^{2+}$ (Ota, 2014, 2017a, 2017b, 2017c, 2017d). Among 53 samples in Acke's list (Acke 2010), 12 samples are Type-B. Typical samples are HD85567 and HD141569. Figure 5 is a comparison with a calculated spectrum of $(C_{23}H_{12})^{2+}$. There are 16 observed peaks in HD85567, which are all reproduced very well by calculation both for wavelength and relative intensity. In observed samples, there are some mixture type spectra. As shown in Figure 6, spectrum of HD142527 (black curve) is well reproduced by a mixture of calculated spectrum of $(C_{23}H_{12})^{2+}$ and $(C_{12}H_8)^{2+}$, which means that such interstellar dust cloud may include both Type-B and Type-D molecules.

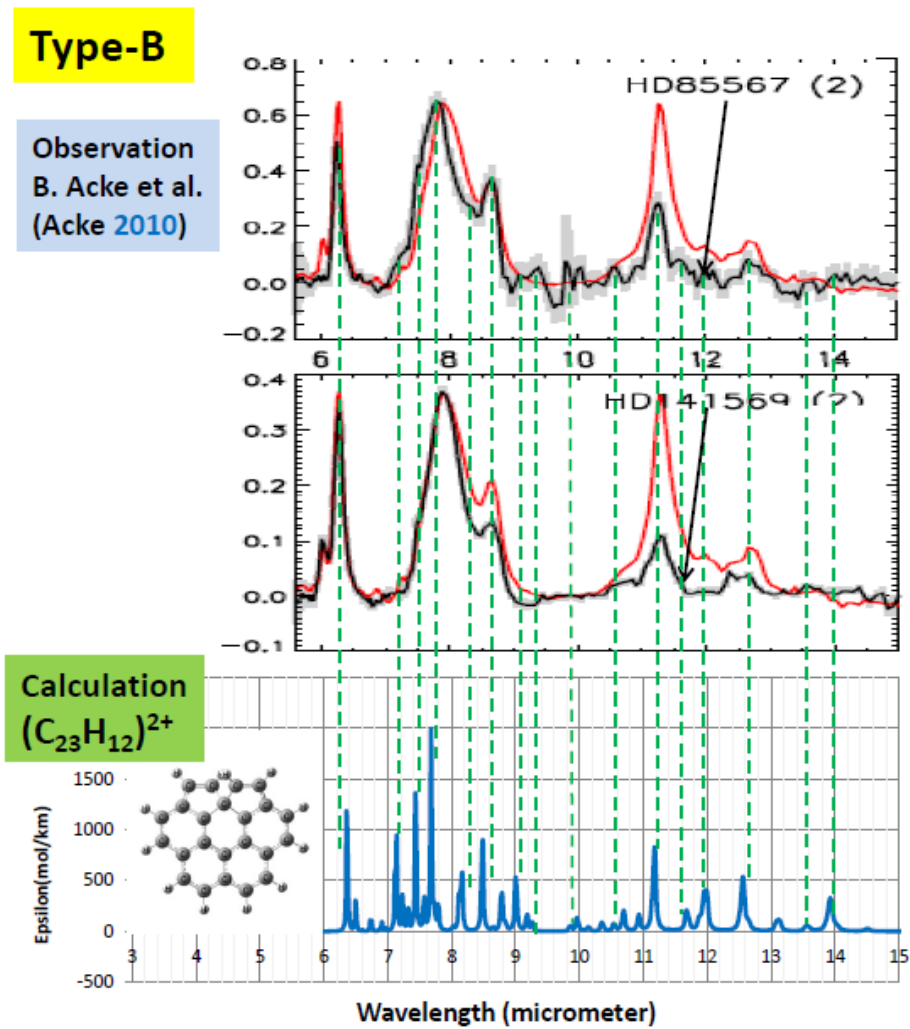

Figure 5, Comparison of observed Type-B spectrum (black curves) and calculated spectrum of $(C_{23}H_{12})^{2+}$ (blue curve). Observed 16 peaks in HD85567 are all reproduced very well by calculated spectrum lines both on wavelength and relative intensity.

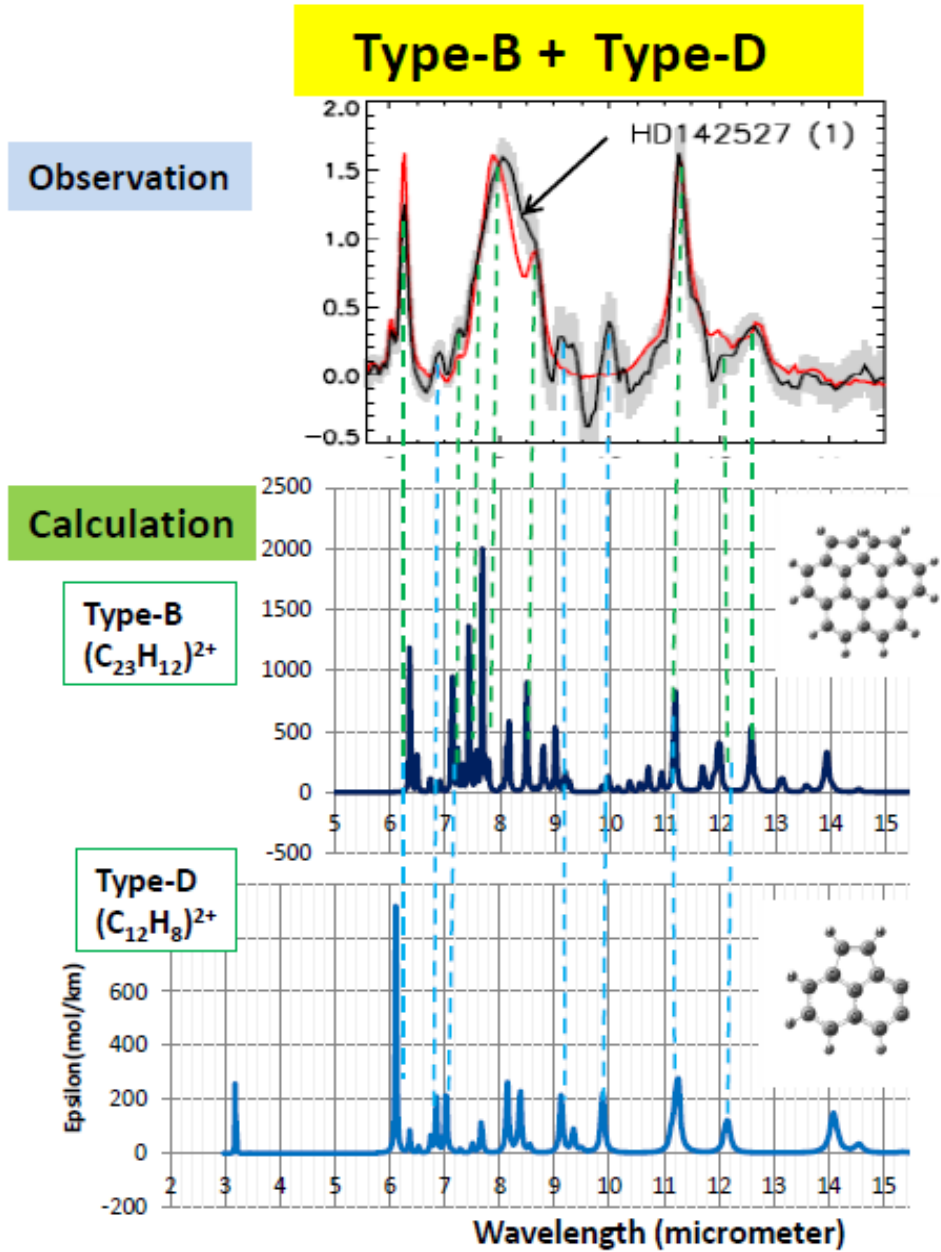

Figure 6, Observed spectrum of HD142527 (black curve) can be explained by a mixture of Type-B and Type-D. Observed spectrum is well reproduced by a suitable mixture with calculated $(C_{23}H_{12})^{2+}$ (dark blue curve) and $(C_{12}H_8)^{2+}$ (blue).

# 7, CONCLUSION

Carrier molecule of infrared spectrum (IR) in Herbig Ae young star was studied by quantum chemical calculation.

(1) Observed IR listed by B. Acke et al. was categorized to four classes, Type-A, -B,-C, and -D.
(2) Among 53 observed samples, there are 26 major samples named Type-D, which shows common IR peaks at 6.2 8.3, 9.2, 10.0, 11.3, 12.1, and 14.0 micrometer. Typical star is HD144432.
(3) Di-cation molecule $(C_{12}H_8)^{2+}$ having hydrocarbon one pentagon combined with two hexagons could reproduce Type-D as like at 6.1, 8.2, 9.2, 9.9, 11.3, 12.2, and 14.1 micrometer.
(4) There are some variations in Type-D. Spectrum of HD37357 was explained by a mixture of di-cation $(C_{12}H_8)^{2+}$ and tri-cation $(C_{12}H_8)^{3+}$.
(5) Ubiquitously observed Type-B was observed in 12 samples of Acke's list. In case of HD85567, observed 16 peaks were precisely reproduced by a single molecule $(C_{23}H_{12})^{2+}$.
(6) There are some mixture case with Type-B and Type-D. Spectrum of HD142527 could be well explained by a mixture of $(C_{23}H_{12})^{2+}$ [Type-B] and $(C_{12}H_8)^{2+}$ [Type-D].

In this study, we could identify carrier molecules for all IR types of Herbig Ae star.

**Submit to arXiv.org. ,     September     , 2017 by Norio Ota**